\documentclass[doublecol]{epl2}
\usepackage{graphicx}
\usepackage{bm}
\usepackage{color}
\usepackage{multirow}
\usepackage{amssymb}

\begin{document}

\title{Nonlinear regime of the mode-coupling instability in 2D plasma crystals}

\author{T. B. R\"ocker\inst{1}\thanks{\email{tbr@mpe.mpg.de}}, L. Cou\"edel\inst{2}, S. K. Zhdanov\inst{1}, V. Nosenko\inst{1,*}, A. V. Ivlev\inst{1}, H. M. Thomas\inst{1,*}, \and G. E. Morfill\inst{1}}

\institute{
  \inst{1}{Max Planck Institute for Extraterrestrial Physics, 85741 Garching, Germany}\\
  \inst{2}{CNRS, Aix-Marseille-Universit\'e, Laboratoire de Physique des Int\'eractions Ioniques et Mol\'eculaires, 13397 Marseille Cedex 20, France}\\
  \inst{*}{Present address: Deutsches Zentrum f\"ur Luft- und Raumfahrt, Forschungsgruppe Komplexe Plasmen, 82230 We{\ss}ling, Germany}
}

\shortauthor{T. B. R\"ocker \etal}


\abstract{
The transition between linear and nonlinear regimes of the mode-coupling instability (MCI) operating in a monolayer plasma crystal is studied. The mode coupling is triggered at the centre of the
crystal and a melting front is formed, which travels through the crystal. At the nonlinear stage, the mode coupling results in synchronisation of the particle motion and the kinetic temperature of the particles grows exponentially.
After melting of the crystalline structure, the mean kinetic energy of the particles continued to grow further, preventing recrystallisation of the melted phase. The effect could not be reproduced in simulations employing a simple point-like wake model. This
shows that at the nonlinear stage of the MCI a heating mechanism is working which was not considered so far.  
}
\maketitle

\section{Introduction}

Complex (dusty) plasmas are weakly ionised gases containing micron-sized dust particles (``microparticles'') which charge up by the absorption of the ambient plasma species (the particle charge $Q$ is roughly a thousand electron charges per micron of diameter).
In the laboratory, complex plasmas are often created by radio-frequency (RF) discharges. Typically, the microparticles are confined in the sheath region where the (averaged) electric field is strong enough to compensate for gravity.
Due to their strong mutual interaction, the microparticles can form strongly coupled (quasi-)two-dimensional (2D) crystalline phases (analogous to colloids \cite{Morfill2009}),
called ``plasma crystals'' \cite{Samsonov2001,Nunomura2000,Schweigert2002,Samsonov2005}. Here, the typical lattice constants $a$ are of the order of several hundred microns. 
The finite vertical extent of such monolayer essentially depends on the strength of the vertical confinement \cite{Samsonov2005}.

The sheath electric field induces a strong ion flow. The microparticles act as ``lenses'' focusing the ion downstream in a close region behind them. This region is vertically extended over a fraction of the effective screening length \cite{Rocker2012a}.
This corresponds to a notable polarisation of the particle screening cloud
 and is often referred to as ``plasma wake'' \cite{Ishihara1997,Lampe2000,Melzer2000a,Hou2001,Vladimirov2003,Samarian2005,Miloch2010,Kompaneets2007}.
The most common way to represent the plasma wake is to add a point-like positive wake charge $q$ at a fixed distance $\delta$ downstream of the particle \cite{Ivlev2001, Zhdanov2009}.
In this ``Yukawa/point-wake'' model, the wake-mediated mutual microparticle interaction is then represented by a superposition of particle-particle and particle-wake interactions both modelled by spherically symmetric Yukawa (Debye-H\"uckel)
potentials with effective screening length $\lambda$.
Although linear response models \cite{Ishihara1997,Kompaneets2007,Rocker2012,Rocker2012a} provide better agreement with experimental results, Yukawa/point-wake model makes the analysis of wake-based effects more transparent.

In 2D plasma crystals, longitudinal and transverse acoustic wave modes can be sustained. The corresponding particle motion is purely horizontal and therefore these modes are usually termed as ``in-plane'' modes.
Due to the finite vertical confinement (with eigenfrequency $f_{\rm v}$), a third fundamental wave mode associated with vertical (``out-of-plane'') oscillations is established.
The out-of-plane mode has an optical dispersion relation \cite{Vladimirov1997,Dubin2000,Ivlev2001,Qiao2003,Samsonov2005,Zhdanov2009}.

The shape of the dispersion relations depends critically on the magnitude of the effective wake dipole moment \cite{Roecker14}:
Under typical experimental conditions it is rather large [$\simeq (0.2$\,-\,$0.3)|Q|a$], which results in an ``attraction effect'' between the in-plane compressional mode and the out-of-plane mode \cite{Roecker14}.
When the vertical confinement is sufficiently weak and the interparticle forces are strong enough, the attraction results in mode coupling giving rise to lower and upper hybrid modes \cite{Roecker14}.
The hybrid modes manifest in the phonon spectra as ``hot spots'' \cite{Couedel2010}. Their intensities in the longitudinal and transverse spectra are equal, due to elliptical polarisation \cite{Zhdanov2009,Roecker14}.  

While the lower hybrid mode is decaying and thus does not affect the stability of the crystal, the growing upper hybrid mode induces the mode-coupling instability (MCI)
when the neutral gas damping is sufficiently weak \cite{Couedel2011}. The onset of this primary {\it plasma-crystal specific} melting scenario can be well explained by linear theory (i.e., assuming small perturbations of the particle positions) based on
Yukawa/point-wake model \cite{Ivlev2001,Zhdanov2009,Couedel2011}. After the onset of MCI, a melting front is formed which propagates in the direction of the maximum instability \cite{Williams2012}. Once the crystal becomes disordered after the mode-coupling onset, the pertubations get large and
thus the aforementioned approach is no longer applicable. A more detailed investigation of this \emph{nonlinear} evolution of the MCI is called for. 

In this Letter, we report investigations concerning the transition between the regimes of linear and nonlinear MCI. During the transition, a stripe pattern in the interparticle separation map 
indicates synchronisation of the microparticle oscillations. The instability continued to grow until complete destruction of the crystal lattice during the nonlinear stage. This is most probably driven by energy absorption from the ion flow 
(involving the spatial variation of the particle charge and/or the screening parameter \cite{Vaulina1999,Zhdanov2005,Kompaneets2006}). The underlying mechanism cannot be explained within the linear regime: Simulations considering the latter
show that the melting might saturate when the monolayer is greatly ``over-heated'' and actually disordered.

\section{Experiment}
\begin{figure}[htbp]
\centering
\includegraphics[width=\columnwidth]{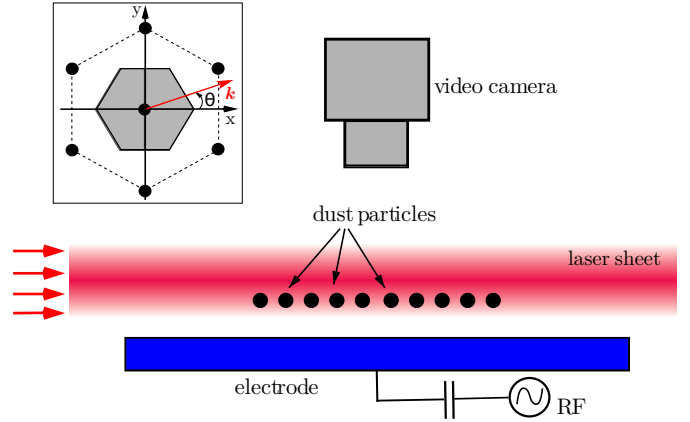}
\caption{\label{fig:schematic}(colour online) Sketch of the experimental setup. The microparticles are confined above the RF electrode and are illuminated
with a horizontal laser sheet having a Gaussian profile in the vertical direction. The inset shows the
elementary cell of hexagonal lattice and the frame of reference chosen in this paper. The orientation of the wave vector
${\bf k}$ is measured in respect to the $x$ axis. The gray region
inside the elementary cell depicts a Voronoi cell.}
\end{figure}

A sketch of the experimental setup is shown in Fig. \ref{fig:schematic}: In a modified version of the Gaseous Electronics Conference (GEC) RF reference cell filled with argon gas (pressure $p=$ 0.5\,-\,1\,Pa),
a capacitively coupled RF glow discharge (13.56 MHz at power $P=$ 5\,-\,20\,W) was established. 
The self-bias voltage was between $-60$~V and $-130$~V. The plasma parameters in the bulk discharge were
deduced from Langmuir probe measurements \cite{Nosenko2009} yielding the electron temperature $T_e=2.5$~eV and the electron density
$n_e=2 \times10^9$~cm$^{-3}$ (at $p=0.66$\,Pa and $P=20$\,W).

A monolayer was formed by levitating melamine-formaldehyde microspheres in the sheath above the RF electrode. The microparticles had diameter $9.19\pm0.14~\mu$m and mass $m=6.1\times10^{-13}$~kg.
The horizontal extent of the crystal was about 50\,-\,60\,mm, depending on the particle number. Here, we report a detailed investigation of the MCI at $p=1.06$\,Pa and $P=20$\,W. 
In this condition, the mean interparticle distance was $a=400\pm20$ $\mu$m in the centre of the monolayer.

The microparticles were illuminated by a horizontal laser sheet which had a Gaussian profile (with standard deviation $\simeq 75~\mu$m) 
in the vertical direction.
The sheet thickness was approximately constant across the whole crystal.
The particles were imaged through a window at the top of the chamber by a Photron FASTCAM 1024 PCI camera (operating at 250 frames per second).
An additional side-view camera was used to verify that our experiments were carried out with a monolayer of particles.

Using a standard individual particle tracking technique \cite{Rogers2007}, the individual particle coordinates ${\bf r}_j = (x_j,y_j,z_j)$ and velocities ${\bf v}_j = (v_{xj},v_{yj},v_{zj})$ were extracted from each frame with sub-pixel resolution.
The velocity fluctuation spectra
were calculated by applying a fast Fourier transform to the particle currents
\begin{equation}\label{}
V_{\{x,y,z\}}(k,t) = \sum_{j=1}^N v_{\{x,y,z \}j} \, e^{i k x_j(t)}, \nonumber
\end{equation}
where $j$ is the particle index, $t$ is time, and $N$ denotes the number of tracked microparticles. The wave vector $\bf k$ (in the horizontal plane) is along the $x$ direction, i.e., ${\bf k} \cdot {\bf r}_j = kx_j$.
Any particular choice of the $x$-axis orientation (with respect to the lattice principal axis) during the analysis allowed the investigation of wave propagation along any direction.
Further on, we indicate the propagation direction by the angle $\theta$, as shown in the inset of Fig. \ref{fig:schematic}. 

Fitting the fluctuation spectra measured at the onset of MCI [$f_{\rm v}=(30\pm1)$\,Hz] by the theoretical dispersion relations \cite{Nunomura2002,Couedel2011,Rocker2012}, 
we evaluated the particle charge $Q=-(20600\pm20\%)e$ ($e$ is the elementary charge) and screening parameter $\kappa=a/\lambda = 0.7$\,-\,$0.95$.
The longitudinal and transverse sound speeds were $C_{\parallel}=36.9\pm5.0$~mm/s and $C_{\perp}=7.3\pm1.5$~mm/s, respectively, calculated from the slopes of the in-plane dispersion relations.

\begin{figure}[t]
\centering
\includegraphics[width=\columnwidth]{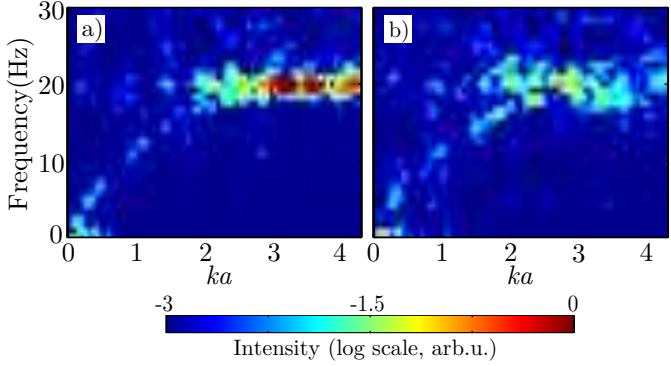}
\caption{(colour online) Longitudinal fluctuation spectra of velocity. The dust lattice waves propagate in a 2D monolayer (the lattice constant is $a=0.40$\,mm) along the principal lattice axes.
Panel a) is for the wave propagation along $\theta=0^{\circ}$, panel b) for $\theta=30^{\circ}$. The intensity (in log scale) is color-coded,
revealing the dispersion of the mode. The hot spots bloom the spectral intensities at smaller and larger $k$. For this example, the argon pressure was $p=1.06$\,Pa and the RF power was $P=20$\,W.}\label{fig:expspectra}
\end{figure}

The emergence of dark-red spots in the longitudinal spectra (Fig. \ref{fig:expspectra}) at the frequency $f_{\rm hm}\simeq20$\,Hz indicates the position of the hybrid mode in the $(k,f)$ plane. The growth of the hybrid mode is very
strong and thus the hot spots bloom the spectral intensities at smaller and larger $k$. The latter are better resolved and more clearly visible \emph{before} the onset of MCI \cite{Couedel2010,Couedel2011}.
Characteristically, the hot spots leak into horizontal energy cascades (toward lower $k$) and are more intense for the propagation direction $\theta=0^{\circ}$ (Fig. \ref{fig:expspectra}a) than for $\theta=30^{\circ}$ (Fig. \ref{fig:expspectra}b),
indicating that $\theta=0^{\circ}$ is the direction of maximum instability growth \cite{Zhdanov2009}.


\begin{figure}[t]
\centering
\includegraphics[width=\columnwidth]{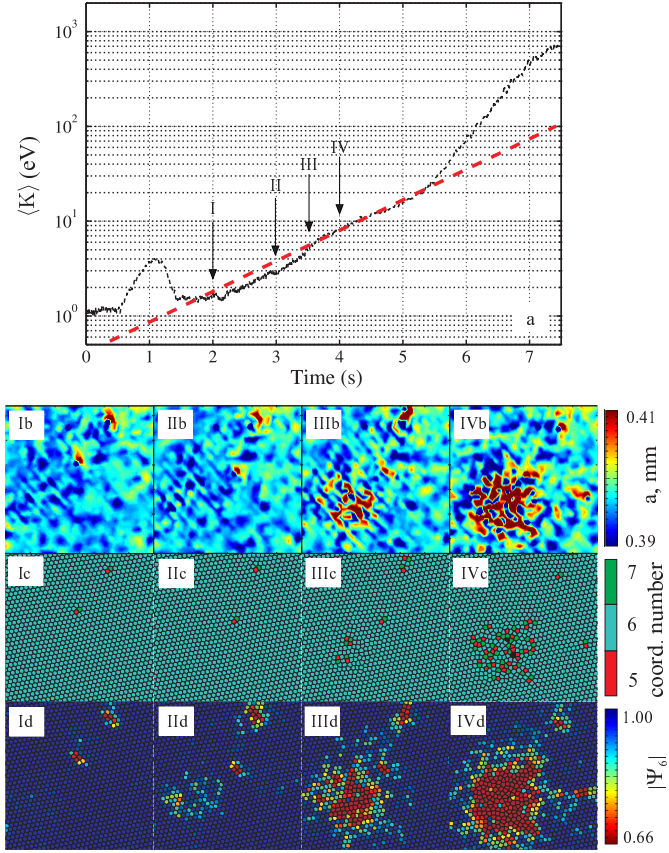}
\caption{(colour online) Mode-coupling induced melting of a 2D monolayer at argon pressure $P=1.06$\,Pa and rf power $P=20$\,W.
 a) The mean dust particle kinetic energy $\langle K \rangle$ as a function of time (the red dashed line is to guide the eyes). For the time interval shwon, the amplitude of the out-of-plane oscillations did not exceed the width of the laser sheet. 
 b) Snapshots of the local interparticle distance $a$ at stages I-IV of the melting.
 c) Snapshots of the corresponding Voronoi cells highlighted by coordination number 5, 6 and 7. Larger and smaller coordination numbers are coloured black.
d) same as (c) with the cells colour-coded by the value of the local bond orientational order parameter $|\Psi_6|$. The timepoints I-IV denote the times 2\,s, 3\,s, 3.5\,s, and 4\,s respectively.}\label{fig:exp}
\end{figure}

In Fig. \ref{fig:exp}, we illustrate the mode-coupling induced melting of a 2D monolayer. Fig. \ref{fig:exp}a is the mean kinetic energy $\langle K \rangle$ of particles as a function of time. Figs. \ref{fig:exp}b-d show the
interparticle separation, Voronoi cells, and the corresponding local bond orientational order parameter, respectively, each measured at four characteristic timepoints labelled by I-IV.  

Let us elaborate on these quantities:
$\langle K \rangle$ is calculated from the in-plane velocities by 
\begin{equation}
  \langle K \rangle = \frac{m}{2N} \sum_{j=1}^N \left[v^2_{xj}(t)+v_{yj}^2(t)\right]. \nonumber
\end{equation}
The local bond-orientational order parameter of the $n$-th particle $|\Psi_6(n)|$ is computed from its bond angles $\theta_{nj}$ with the $N_b$ nearest-neighbour particles by
\begin{equation}\label{eq:eqpsi6}
\Psi_6(n) = \frac{1}{N_b} \sum_{j=1}^{N_b} e^{i6\theta_{nj}},
\end{equation} 
where $\theta_{nj}$ is relative to an arbitrary, but fixed, reference axis.

$|\Psi_6|$ indicates the shear deformation grade of the corresponding Voronoi cell: A perfectly hexagonal Voronoi cell is indicated by $|\Psi_6| = 1$, which decreases
when the cell is deformed. This is highly sensitive to any shear strain in the crystalline structure and thus crystal disturbances are easily detectable, well before the appearance of crystal defects.

At the onset of MCI (timepoint I), the monolayer exhibited a typical hexagonal structure (Fig. \ref{fig:exp}.Ic) \cite{Nosenko2009a}, i.e., $|\Psi_6| \lesssim 1$ almost everywhere (Fig. \ref{fig:exp}.Id).
The interparticle separation was homogeneous \cite{Nosenko2009a} except for a slight stripe pattern and a small number of defects (red spots in Figs. \ref{fig:exp}.Ib-d). 
The instability was triggered in the region of highest particle density (central red spots in Figs. \ref{fig:exp}.Ib-d) and $\langle K \rangle$ started to grow at the rate $\Gamma \simeq 1.7\ \mathrm{s}^{-1}$.

A melting front was formed (timepoint II) which rapidly expanded outward (timepoint III) from the centre toward the edges: The lattice structure got distorted, indicated by
the decrease of $|\Psi_6|$ (Fig. \ref{fig:exp}.IId), and more and more deformed Voronoi cells (Fig. \ref{fig:exp}.IIc).
The particle density became inhomogeneous, which is reflected by the sharpening and elongation of the stripe pattern
in the map of interparticle distances (Fig. \ref{fig:exp}.IIb,c). The stripes' orientation indicates the direction of strongest confinement for which the instability growth is the fastest \cite{Couedel2011}.

Together with the appearance of stripes, the particle motion \emph{synchronises} itself. (Compare movie in \cite{movie}. A detailed study of synchronisation will be published elsewhere.)
This effect was not observed along the other principal lattice axes, as the confinement was not completely symmetric in our experimental device.
During the spread of the instability over the lattice, the kinetic energy $\langle K \rangle$ increased exponentially (timepoints II and III in Fig. \ref{fig:exp}a) with practically constant growth rate.
The rapid growth of $\langle K \rangle$ indicates the fast transition from linear to nonlinear stage of MCI resulting in rather large perturbations of the particle positions.

Shortly after timepoint IV, at which the stripe pattern almost disappeared (Fig. \ref{fig:exp}.IVb), the growth rate increased notably at $\langle K \rangle \simeq 20$\,eV.
Here, $\langle K \rangle$ was practically equal to the hexatic transition temperature $ \approx$ 18\,eV \cite{Nuno06,Quinn01,Knapek07,Ivlevbook}, but below the actual Lindemann melting temperature
$\approx 28$\,eV \cite{Stevens93,Meijer91,Lowen93,Lindemann10}. Note, that the mean kinetic energy does not necessarily reflect the local kinetic energy of particles.  

Two seconds later on, the crystal fully melted.
It is noteworthy, that the melted phase did not recrystallise, unless the RF power and/or gas pressure were increased.
This indicates that the dust particle cloud continuously absorbed energy from the ambient plasma, e.g. by wake effects or spatial variations of charge and screening length \cite{Kompaneets2005}.




\section{Simulations}
We performed Molecular Dynamics (MD) simulations \cite{Donko2009,Zhdanov2012} written in CUDA C language and ran on GPU computing cards (NVIDIA Tesla C1060/C2050).

The Beeman predictor-corrector algorithm \cite{Beeman76} was employed to solve the equations of motion
\begin{equation}\label{langevin}
    m\ddot{\mathbf{r}}_i +m\nu \dot{\mathbf{r}}_i =\sum_{j\neq i}\mathbf{F}_{ij} + \mathbf{L}_{i} + {\bf C}_{i},~i=1,2,...,N.
\end{equation}
for a monolayer comprised of $N =16384$ particles (each with a charge $Q<0$ and mass $m$). 

The monolayer was coupled to a thermostat at temperature $T$. ${\bf L}_i={\bf L}({\bf r}_i)$ denote the Langevin forces which were \emph{on average} balanced by friction (damping rate $\nu$)
and defined by their first two momenta
\begin{equation}
    \langle\mathbf{L}_i(t)\rangle=0,\qquad \langle\mathbf{L}_i(t+\tau)\mathbf{L}_j(t)\rangle = 2\nu
    mT\delta_{ij}\delta(\tau).
\end{equation}
Here, $\delta_{ij}$ is the Kronecker symbol and $\delta(\tau)$ denotes the Dirac delta function.

The vectors ${\bf C}_{i}={\bf C}({\bf r}_i)$ represent the confinement forces from the horizontal and vertical parabolic potential wells with eigenfrequencies $f_{\rm h}=0.19$\,Hz and $f_{\rm v}=$ 23\,Hz,
respectively. Under these conditions, the maximum growth rate of the hybrid mode is larger than the damping rate and thus the MCI is triggered.


For the mutual particle interactions, the Yukawa/point-wake model was employed and the wake of the $i$-th particle resided at ${\bf r}_{{\rm w}_i} = {\bf r}_i - \delta \hat{{\bf e}}_z$.
The resulting forces acting between the $i$- and $j$-th particles read
\begin{eqnarray}\label{yukpw}
    \mathbf{F}_{ij} & =  & -\frac{Q^2}{r_{ij}^2}\exp\Big({-\frac{r_{ij}}{\lambda}}\Big)\Big(1+\frac{r_{ij}}{\lambda}\Big)
    \frac{\mathbf{r}_{ij}}{r_{ij}}\nonumber \\
                    &    & +\frac{|Q|q}{r_{{\rm w}_{ij}}^2} \exp\Big({-\frac{r_{{\rm w}_{ij}}}{\lambda}}\Big)\Big(1+
    \frac{r_{{\rm w}_{ij}}}{\lambda}\Big)\frac{\mathbf{r}_{{\rm w}_{ij}}}{r_{{\rm w}_{ij}}},
\end{eqnarray}
where $\mathbf{ r}_{ij} = \mathbf{ r}_{i}-\mathbf{ r}_{j}$ and $\mathbf{ r}_{{\rm w}_{ij}}=\mathbf{ r}_{{\rm w}_{i}}-\mathbf{ r}_{{\rm w}_{j}}$.
$r_{ij}$ denotes the interparticle distance and $r_{{\rm w}_{ij}}$ is the distance from the particle $i$ to the wake of the $j$-th particle.
The forces (\ref{yukpw}) were accounted for without a cutoff radius.  

\begin{figure}[t]
\centering
\includegraphics[width=\columnwidth]{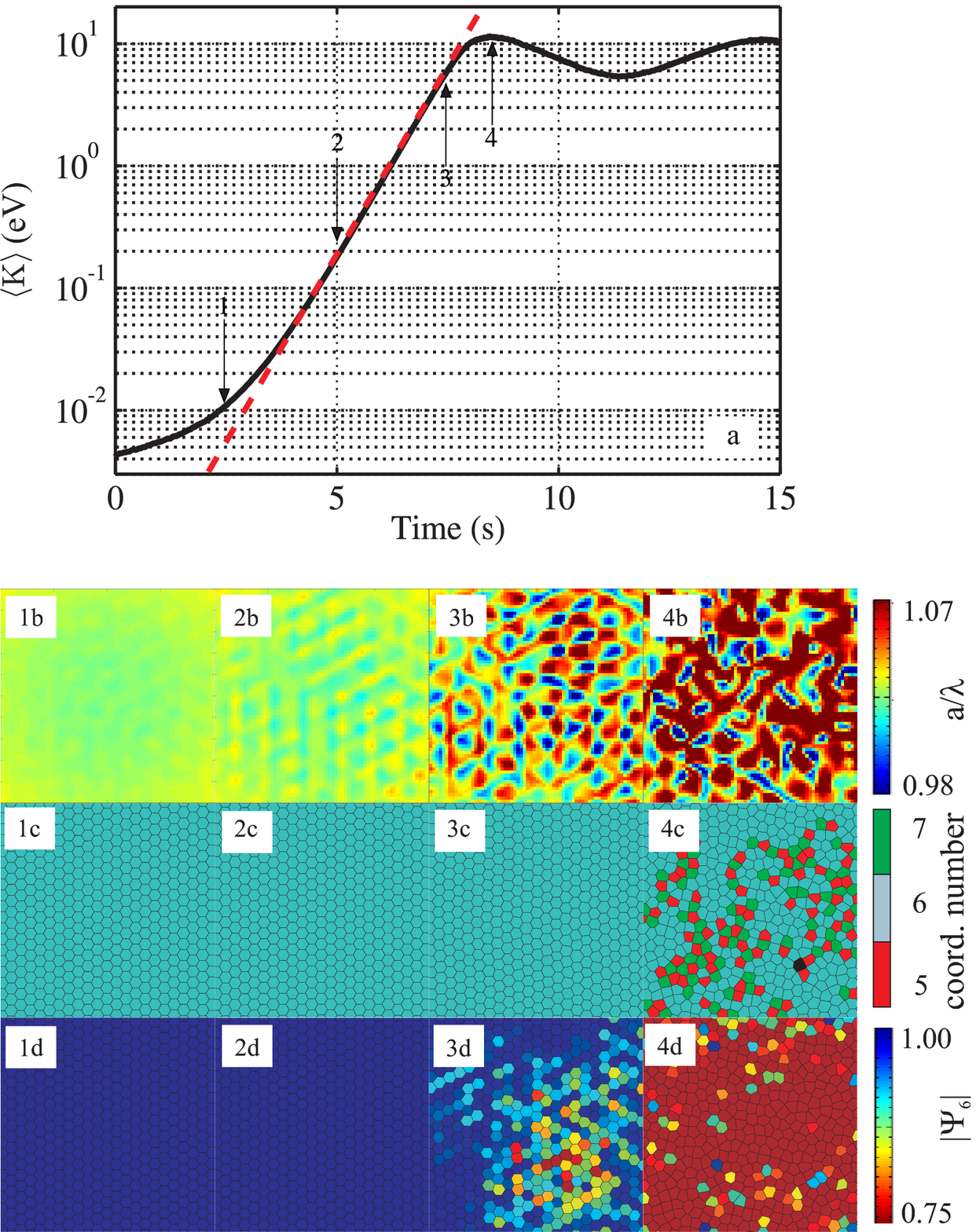}
\caption{(colour online) Same as Fig. \ref{fig:exp}, but the results were obtained from MD simulations: A crystal coupled to a Langevin heat bath and comprised of $N=16384$ particles (mass $m=6.1\cdot10^{-10}$\,g and charge $Q=-19100e$) was simulated.
The confining potential was parabolic in both horizontal and vertical directions with respective eigenfrequencies of $f_{\rm h}=0.19$\,Hz and $f_{\rm v}=23$\,Hz. The effective screening length was $\lambda=400\ \mu$m, the friction rate $\nu=1.48$ s$^{-1}$. 
The timepoints 1-4 denote the times 2.5\,s, 5\,s, 7.5\,s, and 8.5\,s respectively.}\label{fig:sim}
\end{figure}

Fig. \ref{fig:sim} shows the results of our simulations (corresponding to Fig. \ref{fig:exp}):
Shortly after the MCI onset (timepoint 1), the lattice structure showed no defects ($|\Psi_6| \lesssim 1$ everywhere, compare to Fig. \ref{fig:exp}.Ic,d). The crystal was slightly denser in its center, due
to the horizontal confinement forces. The Langevin thermostat induced small thermal fluctuations (Fig. \ref{fig:sim}.1b).

Similarly to experiments, the kinetic energy started to grow nearly exponentially ($\Gamma\simeq0.7\ \mathrm{s}^{-1}$) at timepoint 1. Later on (timepoint 2), the upcoming stripe pattern (in the regions of
highest particle density) indicated the beginning of lattice distortion. The stripes were oriented perpendicularly to the direction of the maximal instability growth, in agreement with our experiments and theoretical predictions \cite{Zhdanov2009}.
Contrary to the experimental counterpart, the deformations of the lattice cells in the simulated crystalline structure (Fig. \ref{fig:sim}.2c,d) were hardly visible whereas the stripe pattern was already observable.

$\langle K \rangle$ reached approximately 6\,eV (timepoint 3) when $|\Psi_6|$ decreased in a notable number of lattice points (Fig. \ref{fig:sim}.3d) at the crystal centre. 
At this point, the stripe pattern in the map of the local interparticle distance was very strong (Fig.~\ref{fig:sim}.3b).


Later on (timepoint 4, $\langle K \rangle \simeq 10$\,eV), the stripe pattern disappeared and the crystal melted. This is indicated by the large number of defects (Fig. \ref{fig:sim}.4c,d).

Most intriguing and unlike experimental observations, the growth of the kinetic energy stopped when the crystal was strongly distorted (Fig. \ref{fig:sim}.4c,d). Finally, $\langle K \rangle$ slowly oscillated around 8\,eV which is notably smaller than the 
kinetic energy measured in the experiments during the highly nonlinear stage (compare Fig. \ref{fig:exp}a after timepoint IV).

The second striking difference in comparison with the experiment is that no propagating melting front was observed. Instead, the lattice defects (their number oscillating with time) were spread over the entire crystal.
In simulation runs where too many structural defects appeared, the kinetic energy stopped growing and even decreased leading to recrystallisation.
When recrystallisation took place, the MCI was retriggered resulting in further cycles of recrystallisation and MCI retriggering. 

It is worth noting that the MCI seems to be highly sensitive to lattice deformations: The kinetic temperature growth
rate was smaller for crystals having such defects. To our best knowledge, this feature was never reported so far. 


\section{Discussion and Conclusion}
The growth of particle kinetic energy during the transition from linear to nonlinear regime of the MCI (observed in both experiments and simulations) is attributed to the exponential growth of the hybrid mode.
The corresponding growth rate $\Gamma_{\rm eff}$ can be calculated from the linear theory: Integration of the instability increment over the highly localised area of the hot spots $S_+$ \cite{Zhdanov2009} and subsequent consideration of the long-time limit yields
\begin{equation}\label{eqth3}
		\Gamma_{\rm eff}(t) \simeq 2 \gamma_{\rm max} - \frac{1}{t} \ln \frac{\sqrt{C_{\parallel}C_{\perp}}S_{+}t}{\pi},
\end{equation}
where $\gamma_{\rm max}$ denotes the maximum instability increment. 

The experimentally observed increase of the energy growth rate during the highly nonlinear stage cannot be explained by the linear theory and thus a mechanism different from the bare growth of the hybrid mode must be responsible. 


To summarise: In experimental observations of the MCI, a stripe pattern in the map of interparticle distance was observed, affecting a quite large area of the crystal (diameter $\simeq10a$).
The stripes indicated regions of shortened and elongated interparticle distances.
At this nonlinear stage, the particle motion turned out to be synchronised. A few seconds later a melting front was formed, which travelled outward from the centre.


The nonlinear energy growth continued after the crystal was melted, which is attributed to the (wake-induced) microparticle energy absorption from the ion flow.
This feature was not observed in simulations of the linear regime, where the MCI-induced lattice distortions were distributed over the whole lattice (instead of forming a melting front) and disturbed the resonant mechanism.
If too many defects appeared, the MCI broke down and the energy growth stopped, followed by recrystallisation.
It can be concluded that in the nonlinear regime of the MCI a heating mechanism is working which cannot be described by the linear theory.
This mechanism was not studied so far and requires rigorous investigation which will be reported in future communications.

\begin{acknowledgments}
We appreciate funding from the European Research Council under the European Unions Seventh Framework Programme (FP7/20072013)/ERC Grant
Agreement 267499 and the French-German PHC PROCOPE program (Project 28444XH/55926142). T.B. R\"ocker acknowledges Alexandra Heimisch for the helpful support.
\end{acknowledgments}

\end{document}